\documentclass[twocolumn]{aastex631}
\usepackage{graphicx,times}  
\usepackage{amssymb,amsmath}
\usepackage{amsfonts}
\usepackage{natbib}
\usepackage{subfigure}

\shorttitle{eMSTO of NGC 6819}
\shortauthors{Yang et al.}
\graphicspath{{./}{figures/}}
\received{Jan 27, 2021}
\accepted{Nov 29, 2021}
\submitjournal{ApJ}

\begin{document}

\title{At what mass are stars braked? The implication from the turnoff morphology of NGC\,6819 }


\email{lichengy5@mail.sysu.edu.cn, yanghuang@ynu.edu.cn, x.liu@ynu.edu.cn}

\author[0000-0002-4028-1893]{Yong Yang}
\affiliation{South-Western Institute for Astronomy Research, Yunnan University, Kunming 650500, P.\,R.\,China}
\affiliation{Department of Astronomy, Yunnan University, Kunming 650500, P.\,R.\,China}

\author[0000-0002-3084-5157]{Chengyuan Li}
\altaffiliation{Corresponding authors}
\affiliation{School of Physics and Astronomy, Sun Yat-sen University, Zhuhai 519082, P.\,R.\,China}

\author[0000-0003-3250-2876]{Yang Huang}
\altaffiliation{Corresponding authors}
\affiliation{South-Western Institute for Astronomy Research, Yunnan University, Kunming 650500, P.\,R.\,China}

\author[0000-0003-1295-2909]{Xiaowei Liu}
\altaffiliation{Corresponding authors}
\affiliation{South-Western Institute for Astronomy Research, Yunnan University, Kunming 650500, P.\,R.\,China}


\begin{abstract}

Extended main-sequence turnoffs apparent in most young and intermediate-age clusters (younger than $\sim$2 Gyr) are known features caused by fast rotating early-type (earlier than F-type) stars. Late-type stars are not fast rotators because their initial angular momenta have been quickly dispersed due to magnetic braking. However, the mass limit below which stars have been magnetically braked has not been well constrained by observation. In this paper, we present an analysis of the eMSTO of NGC\,6819, an open cluster of an intermediate-age  ($\sim$2.5 Gyr), believed to be comparable to the lifetime of stars near the mass limit for magnetic braking. By comparing the observation with synthetic CMDs, we find that NGC\,6819 does not harbor an obvious eMSTO. The morphology of its TO region can be readily explained by a simple stellar population considering the observational uncertainties as well as the differential reddening. In addition, the MSTO stars in NGC\,6819 have very small values of average rotational velocity and dispersion, indicating that they have undergone significant magnetic braking. Combining with results in the literature for clusters of younger ages, our current work suggests that the critical age for the disappearance of eMSTO in star clusters must be shorter but very close to the age of NGC\,6819, and this, in turn, implies a critical stellar mass for magnetic braking at solar metallicity above but close to 1.54\,$M_{\odot}$ based on the PARSEC model. We emphasize that the phenomenon of eMSTO could provide a unique way to constrain the onset mass of magnetic braking. 
\end{abstract}


\section{Introduction} 
Extended main-sequence turnoffs (eMSTOS) driven by fast stellar rotation \citep{2017NatAs...1E.186D, bastian2016young,bastian2018extended, sun2019stellar} are commonly detected features in the color-magnitude diagrams (CMDs) of young and intermediate-age clusters (younger than 2 Gyr) \citep[]{mackey2007double,milone2015multiple}. Stars with the same mass (and composition and age) but different rotational velocities exhibit different colors owing to the effects of gravity darkening \citep{1924MNRAS..84..665V} and rotational mixing \citep{2000ARA&A..38..143M}. Several studies have shown that there is a clear correlation between the stellar rotation and the appearance of an eMSTO \citep{bastian2018extended,2018ApJ...863L..33M,marino2018different,sun2019stellar,2020MNRAS.492.2177K}. 

The CMD morphology of a MSTO region mimics a pattern similar to that of stars with an age spread, which can mislead to the conclusion that the cluster has experienced a continuous star formation history \citep{mackey2007double, Mackey2008Multiple, milone2009multiple, goudfrooij2014extended}. Numerous stellar models have predicted that the eMSTO width (if interpreted as the result of an ``age spread'') should be proportional to cluster age \citep{2013ApJ...776..112Y}, which is also seemingly to have been confirmed by observations \citep{niederhofer2015apparent,bastian2016young,bastian2018extended}. 

Low-mass dwarfs rotate considerably slower than massive ones because of magnetic braking \citep{1962AnAp...25...18S,1987MNRAS.226...57M}. Magnetic braking is only efficient for dwarf stars that are less massive than a critical mass. The convective envelopes of those less massive dwarf stars generate winds that are coupled with the magnetic field, taking away their initial angular momentum. Theoretical stellar evolution modeling usually assumes a critical mass for magnetic braking of  1.7\,$M_\odot$  \citep{2019A&A...627A..24G}. However, \citet{2009Magnetic} have detected magnetic fields in Galactic solar-type stars with masses up to $\sim$ 1.5\,$M_\odot$. The critical mass may therefore range from 1.5 to 1.7\,$M_\odot$, but a more precise value has not yet been determined by observation. The critical mass may also be related to metallicity. \citet{georgy2019disappearance} used two sets of stellar models to investigate the mass at different metallicities and predicted that the mass increases with increasing metallicity. 

The eMSTO observed in young and intermediate-age clusters provides us with an alternative approach to constrain the critical mass for magnetic braking. Clusters older than a critical age are expected to not exhibit an eMSTO because the TO stars in those clusters have already been magnetically braked, making them all slow rotators. Thanks to the high-resolution observations taken by the {\it Hubble Space Telescope} (HST) and the Gaia Space Observatory, a large sample of clusters with eMSTOs in the Milky Way and the Magellanic Clouds have been collected \citep{milone2015multiple, 2017ApJ...844..119L,2018ApJ...863L..33M,cordoni2018extended,2018MNRAS.477.4696M}. However, studies of MSTOs of Galactic clusters with ages of $\sim$2 Gyr have not been reported, which prevents an accurate determination of the critical mass for magnetic braking at solar metallicity. 

The paper aims to detect the critical mass for magnetic braking at solar metallicity. We investigate an intermediate-age open cluster, NGC\,6819, with an age $t \,=\,2.5\pm0.2$ Gyr \citep{2013AJ....146...58J} and metallicity $Z \,= \,0.02\pm0.02$ \citep{leebrown2015spectroscopic}. This cluster thus has TO stars with a mass of 1.54 $\pm$0.05\,$M_{\odot}$ according to the PARSEC model \citep{2012MNRAS.427..127B, Marigo2017A}, which makes it an ideal target to explore the critical mass for magnetic braking. Almost all clusters younger than NGC\,6819 exhibit an eMSTO. We aim to examine whether  NGC\,6819 exhibits additional broadening other than those caused by the observational uncertainties in its MSTO region. This will indicate whether the TO stars in NGC\,6819 have been magnetically braked or not. The paper is structured as follows. We present the data reduction and the main results in section 2. The discussion and a summary of our results are presented in section 3.

\section{Data Reduction and Main Results} 
\begin{figure*}
	\centering
	\includegraphics[width=.95\linewidth]{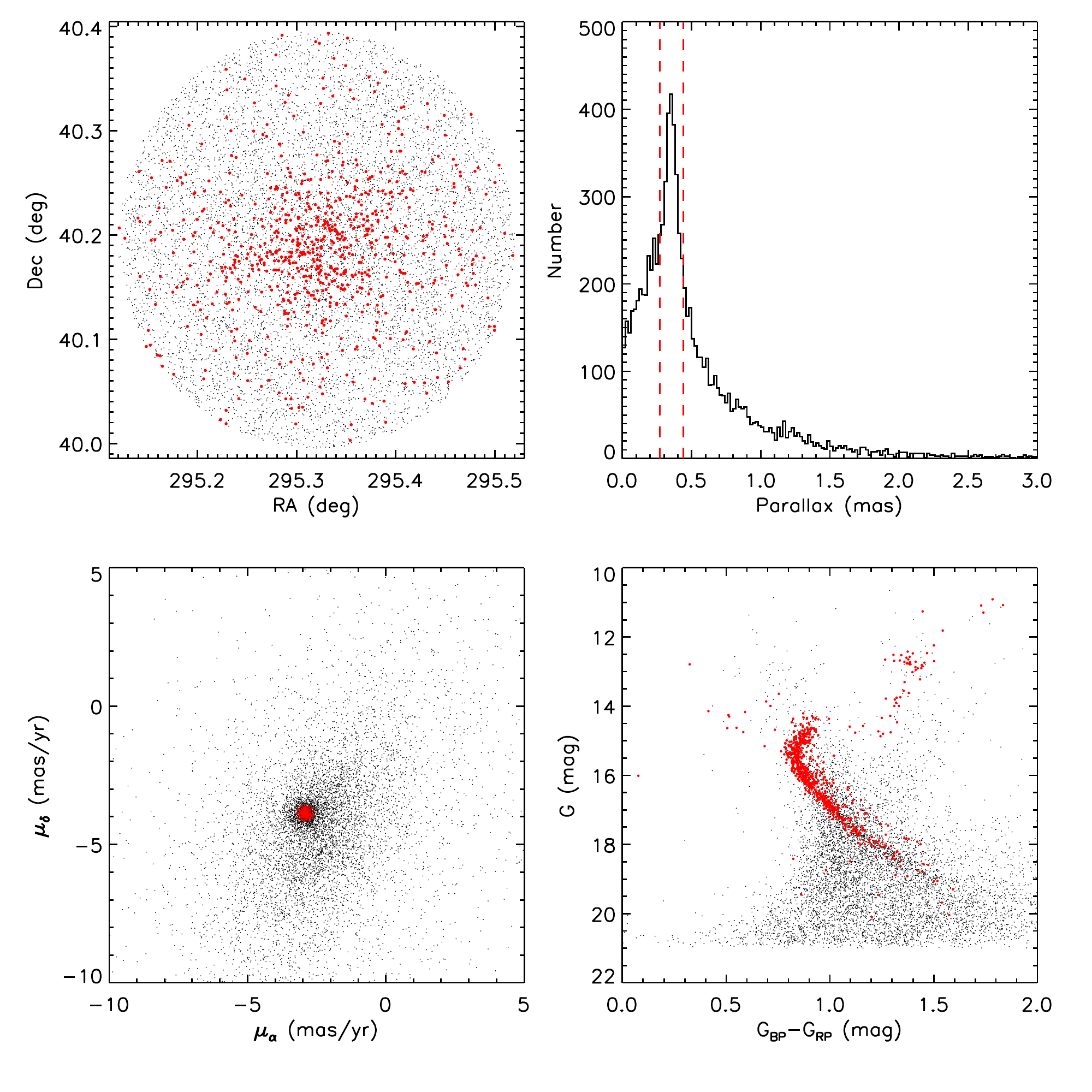}
	\caption{Upper-left: spatial distribution of all stars selected from Gaia DR2 in the field of NGC 6819. Upper-right: the parallax distribution of all stars. The two red dashed lines represent the boundaries applied to select cluster members. Bottom-left: vector-point diagram of the proper motions in the cluster field. Bottom-right: the CMD of all stars in the cluster field. Selected cluster members are represented with red points.}
	\label{fig:member}
\end{figure*}

For the purpose of the current work, we have built a star catalog around the field of NGC\,6819 from the Gaia Data Release 2 \citep[Gaia DR2;][]{2018AA...616A...1G}. NGC$\,$6819 has  central coordinates $\alpha_{\rm J2000}=19^{\rm h}41^{\rm m}16.8^{\rm s}$ and $\delta_{\rm J2000}=40^{\circ}11'42.0''$, and an angular size $r=12'$ \citep{kharchenko2013global}. Stars in the field of view (FoV) exhibit a strong concentration with an average parallax $\pi$=0.355\,mas, indicating a typical distance of 2.82 kpc for NGC\,6819. Using the Gaia DR2 astrometry, we find that the stellar proper motions peak around $<\mu_{\alpha}>$=$-$2.916 mas\,yr$^{-1}$ and  $<\mu_{\delta}>$=$-$3.866 mas\,yr$^{-1}$ due to the common movement of the cluster members. Based on the measured dispersions in the direction of parallax as well as in the proper motion plane, we have filtered all stars in the FoV in the parallax range 0.270 $<\pi<$ 0.441 mas, and within a proper motion difference of $\mid\Delta\mu_{R}\mid=\sqrt{(\mu_{\alpha}-<\mu_{\alpha}>)^2+(\mu_{\delta}-<\mu_{\delta}>)^2}\le$\,0.342 mas\,yr$^{-1}$ with respect to the cluster's bulk proper motions. All those stars are selected as cluster members for the follow-up analysis. The CMD of NGC 6819 after filtering becomes much cleaner than before as shown in  Fig.\,\ref{fig:member}, confirming that most of the field stars have been screened. 

\begin{figure*}
	\centering
	\includegraphics[width=\linewidth]{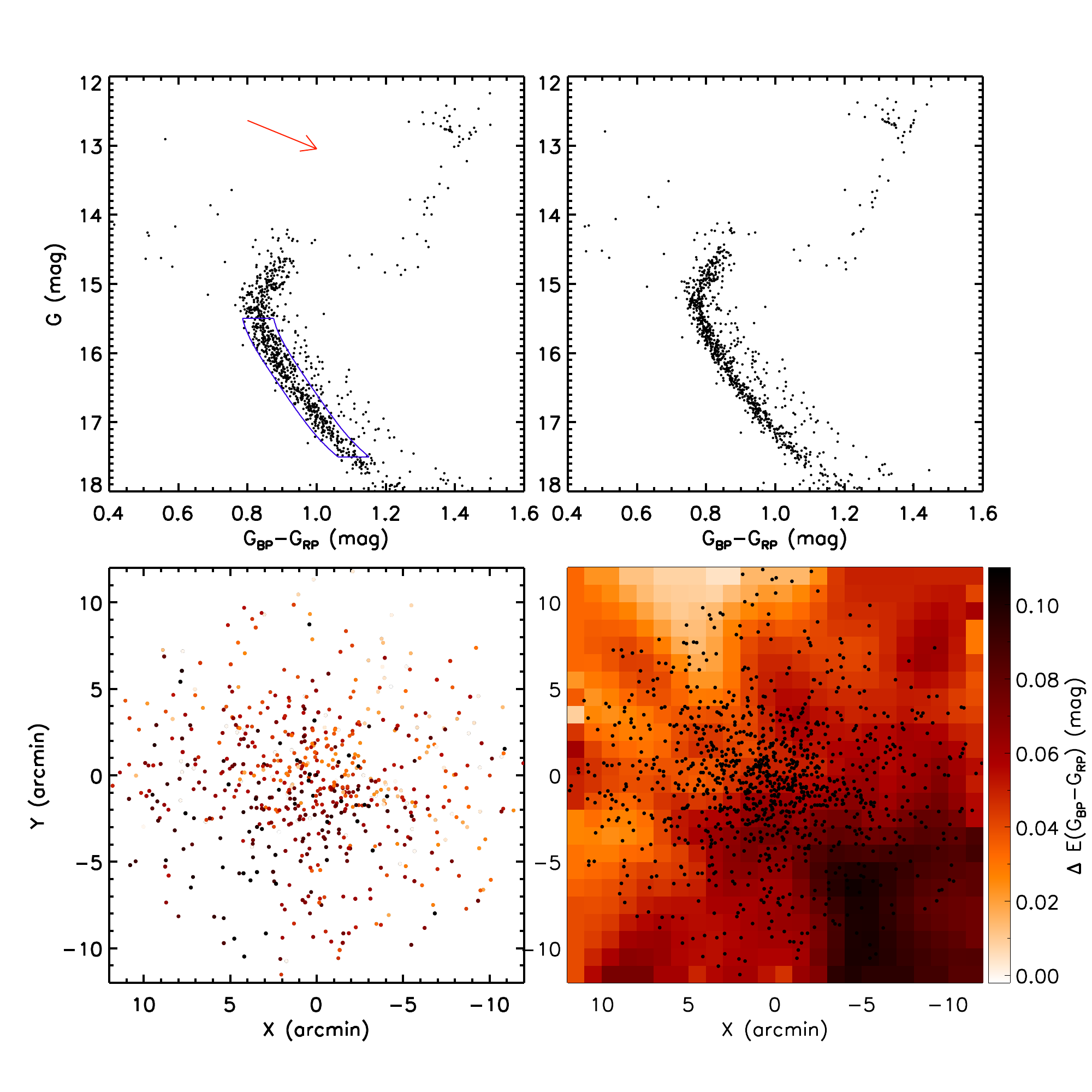}
	\caption{Upper-left: the original CMD of selected cluster members of NGC 6819. The arrow indicates the reddening direction. The blue box shows the stars used to de-redden and their spatial distribution are plotted in the bottom-left panel.  Upper-right: the CMD corrected for differential reddening. Bottom-right: the differential reddening map in the cluster field with all members overplotted. The color indicates the differential reddening value $\Delta E(\textit{G}_{\rm BP}-\textit{G}_{\rm RP})$. In the corners of this field the estimates are extrapolated due to the lack of cluster members.}
	\label{fig:ebv}
\end{figure*}

To examine whether the observed MSTO region is extended or not, we first correct for the differential reddening across the entire FoV using the method developed by \citet{2012ApJ...751L...8P, 2013AJ....146...43P}. An empirical blue envelope of the main sequence is plotted by shifting the MS fiducial line in color. Each star is then de-reddened to match the blue envelope along the reddening direction. Only the upper MS members in the magnitude range of 15.5--17.5 mag are adopted in this step to avoid the influence of photometric errors. The differential reddening value of stars bluer than this envelope is taken as 0, and the stars whose value exceeds 0.09 are excluded because they are more likely to be binary stars. The spatial distribution of the differential reddening value for adopted members is shown in the bottom-left panel of Fig.\,\ref{fig:ebv}. The differential reddening map in the cluster field is divided into grids with a diameter of 1 arcmin, and the differential reddening value of each grid is evaluated according to the nearest six measurements (or all measurements when the number exceeds six). There is a notable reddening gradient that roughly increases from the upper left to the lower right, which is consistent with the reddening map extracted from the NASA/IPAC infrared science archive\footnote{https://irsa.ipac.caltech.edu/applications/DUST/} \citep{1998ApJ...500..525S}. The reddening value of each member star is then estimated by matching its coordinate with the map. The CMD after this differential reddening correction is shown in the upper-right panel of Fig.\,\ref{fig:ebv}, which shows that differential reddening is a dominant contributor to the eMSTO. 

\begin{figure}
	\centering
	\includegraphics[width=\linewidth]{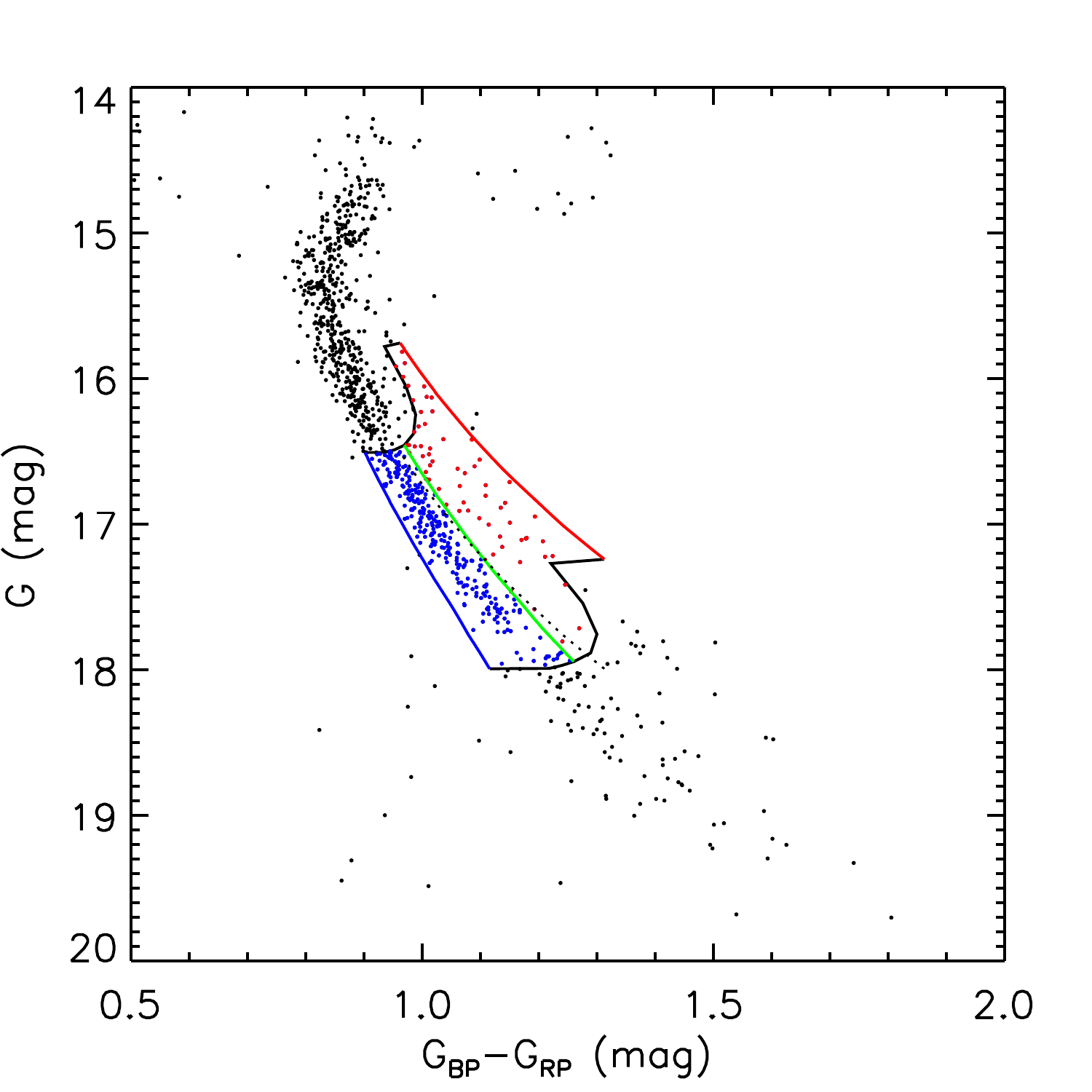}
	\caption{Adopted binary- and single-star regions along the MS for NGC 6819; see text for a detailed description of our approach. }
	\label{fig:bsplot}
\end{figure}

Unresolved binaries are redder and brighter than single MS stars with similar masses and tend to broaden the MS and complicate the morphology of the MSTO. To study the effect of binaries on the broadened MSTO, we first determine the binary fraction along the MS using a method identical to that adopted in \citet{2012A&A...540A..16M}. The approach involves counting the star numbers in the binary area of the CMD. The green line in Fig.\,\ref{fig:bsplot} represents the MS fiducial line shifted by three times the photometric error toward the red in color, which basically coincides with the line of q = 0.6, where q is the mass ratio of the two components of the binary. It is thus difficult to distinguish between binary stars and single stars in the region of q $<$ 0.6 owing to the photometric errors and mixture of single and binary stars. We, therefore, selected the region of q $>$ 0.6 to count the number of binaries. In the faintest section of the MS, the photometric error is so large that there is single-star contamination in the binary region. We, therefore, choose the stars between 16.5 and 18.0 mag in the \textit{G} band to estimate the binary star fraction. We have divided the CMD along the MS into two regions, A and B, which are represented in Fig. 3 as blue and red dots, respectively.  Region A includes all of the single stars and binary systems with q $<$ 0.6, and region B is populated by binaries with q $>$ 0.6. The fraction of binaries is then calculated as $f_{bin}^{q>0.6}={N_{b}}/{N_{all}}$, where $ N_{b}$ is the number of the members in region B and $ N_{all}$ is the number of all members in regions A and B. We assume that q follows a uniform distribution \citep[e.g. ][]{2012A&A...540A..16M, 2016MNRAS.455.3009M, cordoni2018extended}, thus the total binary fraction is $f_{bin}^{tot} \sim 2.5 \times f_{bin}^{q>0.6} \sim 45\%$.

\begin{figure*}
	\centering
	\includegraphics[width=0.95\linewidth]{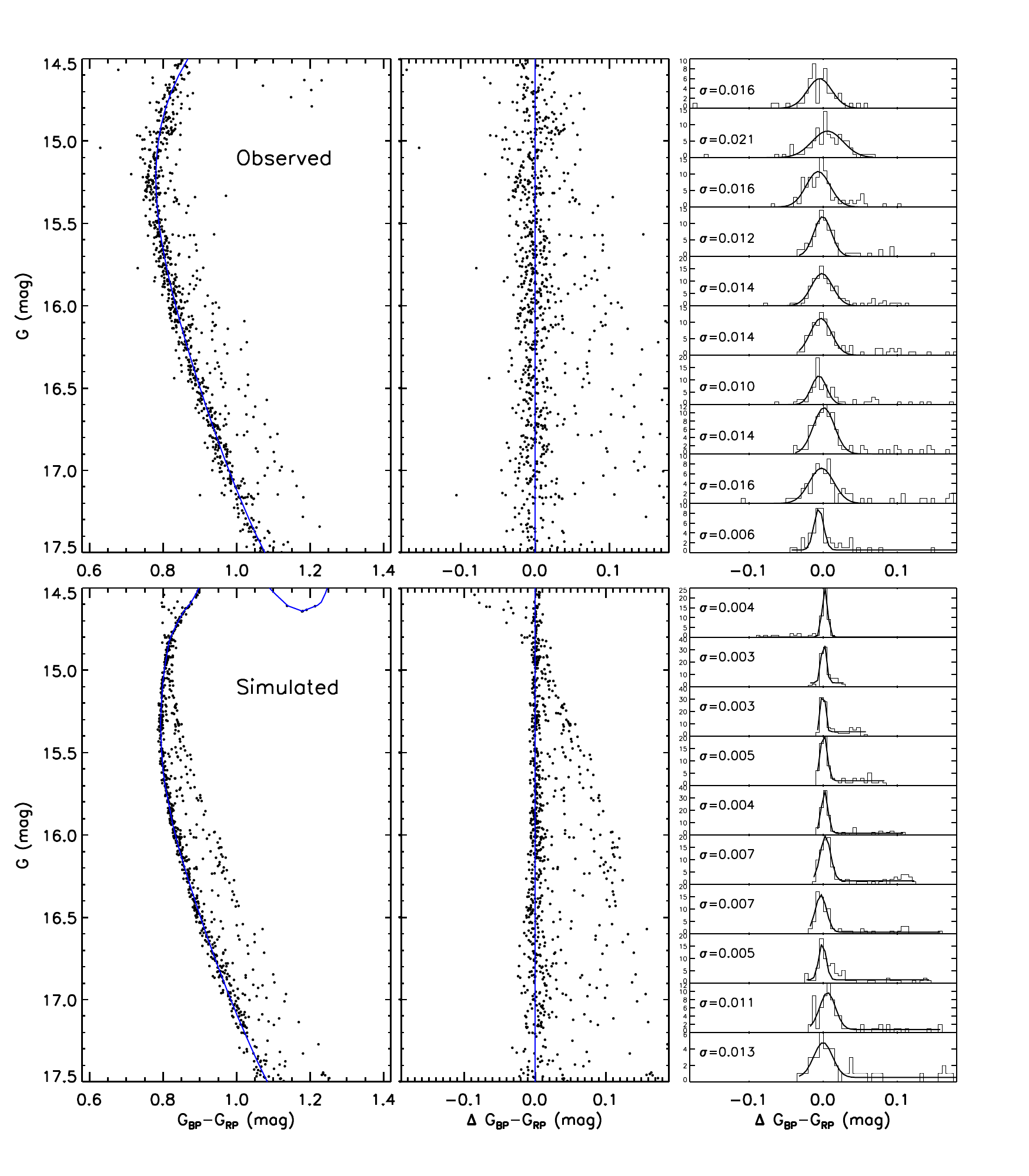}
	\caption{Left: the observed (top) and simulated (bottom) CMD of NGC 6819 with the fiducial line overplotted; Middle: the CMDs rectified by subtraction of the fiducial line; Right: color distribution of the rectified CMDs. The $\sigma$ in the inset is the dispersion of the best-fitting Gaussian. }
	\label{fig:modebv}
\end{figure*}

We simulated a synthetic CMD of a simple stellar population (SSP) with the same photometric uncertainties as the real observations in the Gaia \textit{G}, \textit{G}$_{\rm BP}$ and \textit{G}$_{\rm RP}$  passbands, and the same unresolved binary fraction derived above. The simulated CMD is based on the best-fit isochrone derived from the PARSEC model \citep{2012MNRAS.427..127B, Marigo2017A}. The input fit parameters are $\log{(t/{\rm yr})}$ = 9.40 ($\sim$2.5 Gyr), [Fe/H] = 0.0, $(m-M)_0$ = 12.3  mag, and E$(\textit{G}_{\rm BP}-\textit{G}_{\rm RP})$ = 0.21 mag. The number of stars and their luminosities in the synthetic CMD are identical to the observation. A comparison of the lower-part of the simulated MS and the observation shows that the synthetic MS is slightly narrower than that observed by 0.011 mag in color, as shown in Fig.\ref{fig:modebv}. Because stellar rotation does not affect those bottom-MS stars, the additional broadening of the MS must come from the residual differential reddening or some unaccounted for photometric processing effects (e.g., calibration). We directly correct for this difference by adding some additional noise to all of the simulated stars. Fig.\ref{fig:modebv} also compares the width between the MS and MSTO of the observed CMD, which demonstrates a lack of significant spread in color space, while the MSTO width is slightly larger as binaries are included there.

\begin{figure*}
	\centering
	\includegraphics[width=\linewidth]{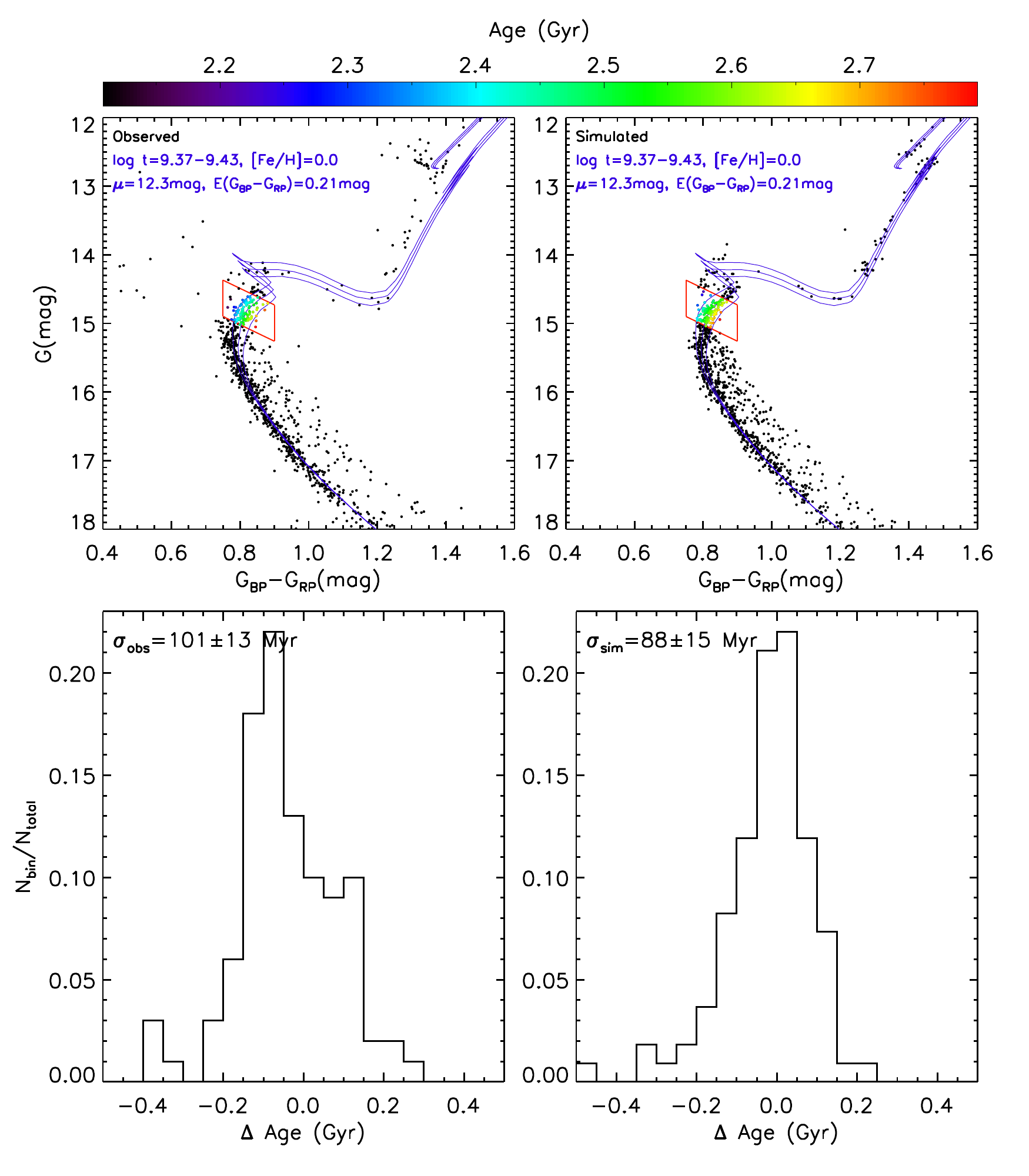}
	\caption{Top-left: the observed CMD of NGC 6819 with field contamination and differential reddening corrected for. Blue solid lines are isochrones of ages $\log{(t/{\rm yr})}$=9.37, 9.40, and 9.42, respectively, that delineate the boundaries and the ridge-line of the MSTO region. Other best-fitting parameters are also marked in the panel. Top-right: the same but for the synthetic SSP. Bottom-left and right: the distributions of the pseudo age differences of TO stars corresponding to the top two panels.}
	\label{fig:obsimcomp}
\end{figure*}

We compare the observed and simulated CMDs to examine whether the observation is consistent with a SSP. The observed and simulated CMDs are presented in the top-left and -right panels of Fig.\,\ref{fig:obsimcomp}, respectively, and the TO region in each CMD is highlighted by a parallelogram. We aim to examine whether the observed width of the TO region is consistent with that of the SSP. To quantitatively compare their widths, we assume that the extension of the MSTO is interpreted as an age spread. We then apply the same method as that adopted in \citet{li2014not}, \citet{Bastian2016A}, \citet{bastian2018extended}, \citet{cordoni2018extended} to the selected TO stars. We generate a grid of PARSEC isochrones in the range of $\log{(t/{\rm yr})}=9.28,...,9.48$, with a step of 0.0005. In addition to the age, these isochrones have other parameters (metallicity, distance modulus, and extinction) that are all identical to the best-fit isochrone. For each selected TO star, we assign the age of the closest isochrone as its best-fit age, as color-coded in Fig.\,\ref{fig:obsimcomp}. We then calculate the ``age distribution'' of both the observation and simulation, and the results are presented in the bottom corresponding panels. We remind readers that the synthetic TO stars do not differ in age, but rather the observed apparent age difference is mimicked by a combination of photometric uncertainties, differential reddening, and unresolved binaries. 

We adopt the same approach used by \citet{milone2015multiple}, \cite{cordoni2018extended} to calculate the age spread of NGC\,6819 to directly compare with their results. In this work, we interpret the observed width of the MSTO region caused by an ``age spread''. We call this age spread a pseudo age spread because it may be simply caused by a combination of photometric uncertainties, differential reddening residuals, and differential rotations.  Our calculation yields an observed pseudo age spread of $\sigma_{\rm obs} = 101\pm13$ Myr. For the simulated SSP, we ran one hundred simulations and took the median as the final result, which is $\sigma_{\rm sim}=88\pm15$ Myr. The two values are consistent within the 1$\sigma$ confidence interval. Actually, the 13-Myr difference can be fully explained by the contamination of blue straggler stars and remaining field stars. We conclude a negligible age difference between the observation and simulation. The observed MSTO width is identical to that of a SSP.  

\begin{figure*}
	\centering
	\includegraphics[width=\linewidth]{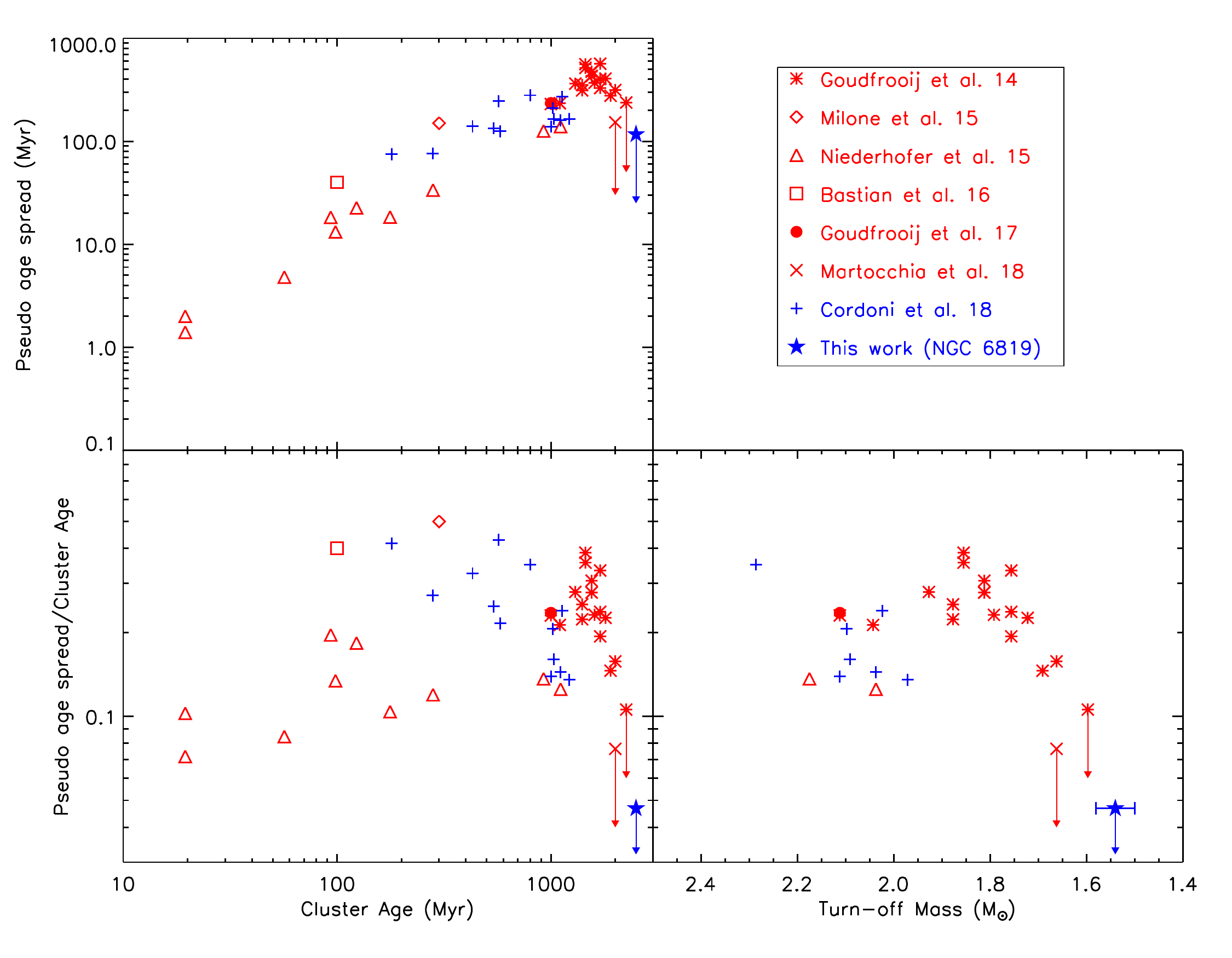}
	\caption{Top-left: the relation between cluster age and pseudo age spread for clusters with eMSTOs. Blue symbols denote Galactic open clusters \citep{georgy2019disappearance}, while red ones are clusters in the Magellanic Clouds. Different symbols represent results from different studies \citep{goudfrooij2014extended, milone2015multiple, niederhofer2015apparent, bastian2016young, 2017ApJ...846...22G, 2018MNRAS.477.4696M}. NGC\,6819 is marked by a blue-filled star. An arrow indicates an upper limit of the pseudo age spread. Bottom-left: the normalized pseudo age spread as a function of cluster age. Bottom-right: the same as the bottom-left panel but as a function of the mass of TO stars. An error bar of the TO mass of NGC 6819 is also drawn.}
	\label{fig:agefwhm}
\end{figure*}

To directly compare our result with previous studies, we calculate the ``intrinsic'' age dispersion of NGC\,6819 and find $FWHM = 2.3548\times\sqrt{\sigma_{\rm obs}^2-\sigma_{\rm sim}^2}\approx117$ Myr \citep{milone2015multiple,cordoni2018extended}. In the top panel of Fig.$\,$\ref{fig:agefwhm}, we present the correlation between the eMSTO width in units of pseudo age spread and the cluster ages for all of the studied Galactic and Magellanic Cloud clusters. The pseudo age spread exhibits an increasing trend from the extremely young clusters ($\sim$20 Myr) to intermediate-age clusters ($\sim$1--2 Gyr), reaches a maximum around a cluster age of $\sim$1.5 Gyr, and then declines with increasing age. Our result is plotted as a blue star. The fact that young clusters show smaller pseudo age spreads does not indicate that rotation is not important for their member stars. Because massive stars evolve considerably faster than low-mass stars, a very small age spread in the TO regions of young clusters yields a much larger color dispersion than in the case of old clusters. The effect can hardly be mimicked by stellar differential rotation. In contrast, for older clusters, a modest color variation caused by differential stellar rotation can mimic a dramatic age spread (up to $\sim$600 Myr). 

\begin{figure}
	\centering
	\includegraphics[width=\linewidth]{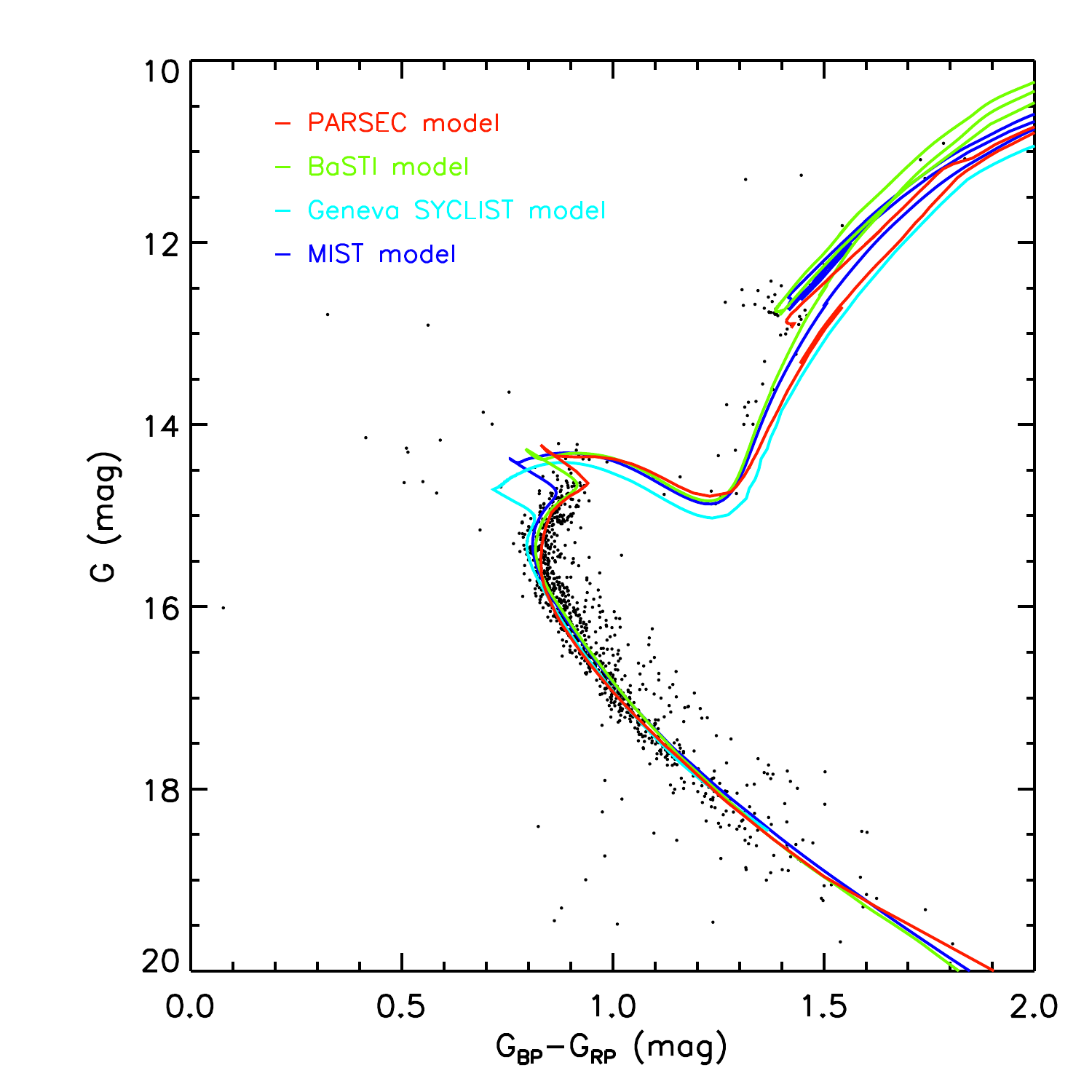}
	\caption{To fit the CMD of NGC\,6819 using four different sets of theoretical stellar evolution isochrones. The red, green, cyan and blue isochrones are PARSEC \citep{2012MNRAS.427..127B, Marigo2017A}, BaSTI  \citep{2004ApJ...612..168P}, Geneva SYCLIST \citep{2014A&A...566A..21G}, and MIST \citep{2016ApJ...823..102C}, respectively.}
	\label{fig:isocomp}
\end{figure}

To minimize the effect of stellar evolution, the bottom-left panel of Fig.\,\ref{fig:agefwhm} shows the pseudo age spread normalized by the real cluster ages as a function of the real cluster ages. We find that the normalized pseudo age spread ranges from $\sim$10\% to 50\% of the cluster age for clusters younger than $\sim$1.5 Gyr. For NGC\,6819, our current analysis yields a value of only 5\%, which indicates that the effect of stellar rotation is negligible in this particular cluster. The downward arrow indicates that the derived intrinsic age spread is an upper limit because the observed width is fully consistent with a SSP after considering the uncertainties. 
As is shown in the Fig.\,\ref{fig:isocomp}, we use four different sets of theoretical stellar evolution isochrones with the same age (2.5 Gyr) and metallicity ([Fe/H] = 0.0) as the input parameters to fit the CMD of NGC\,6819. The four isochrone sets are PARSEC \citep{2012MNRAS.427..127B, Marigo2017A}, MIST \citep{2016ApJ...823..102C}, Geneva SYCLIST \citep{2014A&A...566A..21G} and BaSTI \citep{2004ApJ...612..168P}. They interpret the TO mass of NGC\,6819 as 1.54$\,M_\odot$, 1.46$\,M_\odot$, 1.43$\,M_\odot$, and 1.50$\,M_\odot$, respectively. The models have different underlying assumptions that result in deviations in the derived masses. But the trend of pseudo age spread with TO masses of star clusters does not change. We, therefore, want to discuss the mass range of magnetic braking based on a certain model. A visual inspection of Fig.\,\ref{fig:isocomp} reveals that the PARSEC isochrone has a better fit. We, therefore, select the PARSEC model to re-measure the TO masses of star clusters and assume they're all solar metallicity to determine the mass range for magnetic braking. In the bottom-right panel of Fig.\,\ref{fig:agefwhm}, we convert the clusters' ages into masses of their TO stars. Since the age of NGC\,6819 is about $2.5\pm0.2$ Gyr \citep{2013AJ....146...58J}, we have determined a mass of $1.54\pm 0.04 \,M_\odot$ for its TO stars at solar metallicity using the PARSEC model, which is also represented in Fig.\,\ref{fig:agefwhm}.

\begin{figure*}
	\centering
	\includegraphics[width=\linewidth]{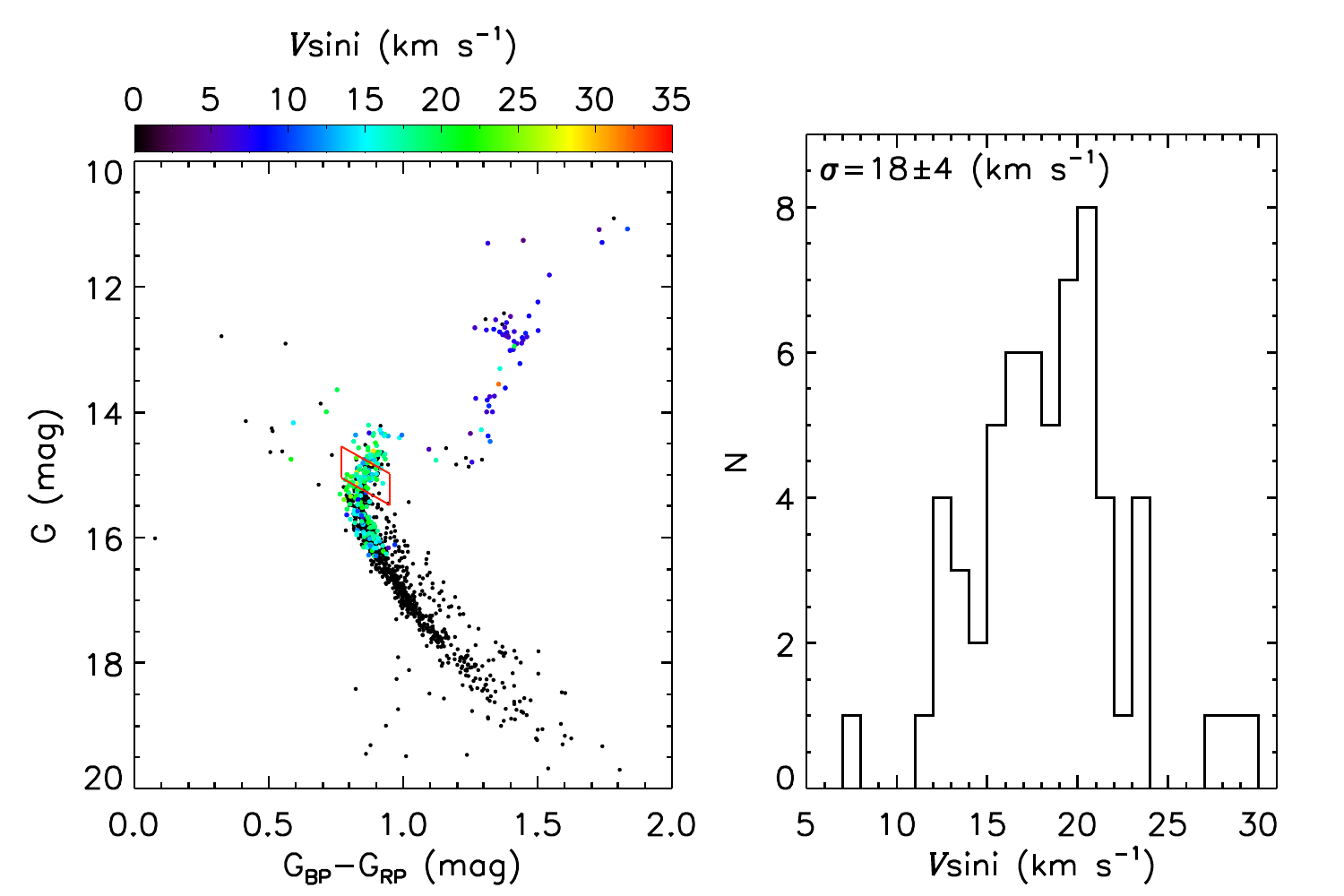}
	\caption{Left: the observed CMD of NGC 6819 with the projected rotational velocities of some stars are color-coded. Right: distribution of projected rotational velocities of TO stars (selected from the red parallelogram in the CMD, as shown in the left panel).}
	\label{fig:wiyncmd}
\end{figure*}

Several previous studies have shown that the eMSTO regions are linked to TO stars of large rotational velocities \citep{bastian2018extended,2018ApJ...863L..33M,marino2018different, sun2019stellar}. Because NGC\,6819 does not exhibit any signature of an eMSTO region, measurements of the rotational velocities of its TO stars can provide crucial information regarding whether its TO stars have been magnetically braked. We thus collected the projected rotational velocities ($V\sin{i}$) of the cluster member stars from \citet{leebrown2015spectroscopic}, who estimated the rotational velocities from the line widths of high-dispersion spectra obtained using the WIYN 3.5-m telescope. Their measurements include TO, sub-giant, red-giant, and red-clump stars. The left panel of Fig.\,\ref{fig:wiyncmd} presents the observed CMD of NGC\,6819, in which the stars are color-coded by their measured $V\sin{i}$ values. The right panel of Fig.\,\ref{fig:wiyncmd} presents the $V\sin{i}$ distribution of the TO stars. The latter yields an average value of $\left<{V\sin{i}}\right>$=$18\pm4$ km$\,$s$^{-1}$. The mean and dispersion of the low-$V\sin{i}$ values of those TO stars are in sharp contrast with that found for younger clusters, where the projected rotational velocities of their eMSTO stars range from $\le\,$50 km$\,$s$^{-1}$ to $\ge\,$300 km$\,$s$^{-1}$ \citep{marino2018different,sun2019stellar}. 

\section{Scientific Implications}
Since the eMSTO region is an indicator of stellar rotation and the rotation speed is a function of stellar mass owing to magnetic braking, one may expect that it is the cluster age that dictates the disappearance of the eMSTO region of a cluster \citep{georgy2019disappearance}. Other than NGC\,6819, two additional clusters,  Hodge\,6 \citep{goudfrooij2014extended} and NGC\,1978 \citep{2018MNRAS.477.4696M}, have been reported to have a narrow TO region. However, those two are Large Magellanic Cloud (LMC) clusters that have lower metallicities than Galactic ones. More importantly, information regarding the rotational velocities of their TO stars is lacking. The distances of Hodge\,6 and NGC\,1978 (about 50 kpc) are roughly 18 times that of NGC\,6819 (2.8 kpc), which introduces larger photometric uncertainties for their TO stars. We thus do not know whether the ``intrinsic'' age spread still indicates the presence of some degree of differential rotation. 

NGC\,6819 sets the strongest constraint on the critical mass for magnetic braking, and exhibits the lowest ``intrinsic'' and normalized age spread amongst all of the clusters (Fig.\,\ref{fig:agefwhm}). As described earlier, stars with a large ``age difference'' only constitute a minor fraction, and can easily be explained by contaminating field stars or blue straggler stars. If one can fully exclude those contaminations, the width of its TO region would be fully consistent with a SSP. Direct $V\sin{i}$ measurements of its TO stars further support the idea that those stars have been magnetically braked. The $V\sin{i}$ values have an average of only 18 km$\,$s$^{-1}$. In comparison, for clusters with eMSTO regions, the rotational velocities of their TO stars can easily reach $\ge$ 300 km$\,$s$^{-1}$  \citep{2019ApJ...876...65L}. 

The very narrow distribution of rotational velocities of the TO stars in NGC\,6819 indicates a pattern shaped by magnetic braking as well. Unlike massive stars, low-mass stars will steadily lose angular momentum via magnetically channeled winds. Any initial dispersion of their rotations will therefore rapidly decrease. As a result, their rotation rates will converge to a certain value that depends on mass \citep{2015Natur.517..589M}. 

Our result thus yields a lower mass limit for magnetic braking of $M_{\rm mb}$=$1.54\pm 0.04\,M_{\odot}$  according to the PARSEC model \citep{2012MNRAS.427..127B, Marigo2017A}. Stars less massive than this critical value should have all been magnetically braked. \citet{goudfrooij2014extended} showed that Hodge\,6 (2.25 Gyr-old), the oldest cluster in their analysis, exhibits a MSTO width that corresponds to an intrinsic age spread of 238\,Myr, which is about double that of NGC\,6819. If the intrinsic age spread of Hodge\,6 is caused by stellar rotation, this would indicate that the critical mass for magnetic braking must lie somewhere in the mass range defined by Hodge\,6 and NGC\,6819. On the basis of the PARSEC models, we determine this mass range as 1.54--1.60\,$M_{\odot}$. 

However, it is also possible that the MSTO region of Hodge\,6 is consistent with a SSP, and its intrinsic age spread is caused by photometric uncertainties rather than stellar rotations. In that case, the critical mass for magnetic braking should be above 1.60\,$M_{\odot}$. Furthermore, as concluded by \citet{georgy2019disappearance}, the mass for magnetic braking may depend on metallicity, in which the onset mass of magnetic braking would increase at higher metallicities. Because Galactic clusters have higher metallicities than Magellanic Clouds clusters (with sub-solar metallicities), the onset mass for magnetic braking should also be higher. In that case, we must compare our result with Galactic clusters rather than Magellanic Clouds clusters. Unfortunately, the oldest Galactic cluster  \citep[1.22 Gyr;][Melotte\,71 ;]{cordoni2018extended} with a known eMSTO region is too young to constrain the critical mass for magnetic braking. This cluster is Melotte\,71 \citep[1.22 Gyr;][]{cordoni2018extended}. If one considers only Galactic clusters, the critical mass for magnetic braking should lie between 1.54 and 1.97\,$M_{\odot}$. More older clusters with an eMSTO should be studied to further tighten the constrain on this mass range, which will be explored in our future work.

\citet{2018ApJ...864L...3G} also proposed that there is a \textit{kink} in the main sequence of young star clusters in the LMC. This \textit{kink} represents the disappearance of rotational effects in stars fainter than the \textit{kink} magnitude, corresponding to initial stellar masses of 1.45 $\pm$0.02\,$M_{\odot}$ based on the PARSEC model. Compared to the mass limit, our value is larger since it's derived at higher metallicity. We adopted the age 2.5 Gyr and LMC metallicity (Z = 0.008) as the input parameters of the PARSEC isochrone, and obtain a TO mass of 1.44\,$M_{\odot}$, which is roughly consistent with that estimated in \citet{2018ApJ...864L...3G}. \citet{2020MNRAS.492.2177K} raised a query regarding the \textit{kink} because they detected fast-rotating stars well below this limit in the LMC cluster NGC 1846. They found that over a range of masses, the fraction of rapid rotators continuously decreases and eventually leaves only the slowly rotating branch. This seems to indicate that there is no critical mass limit below which stars have been magnetically braked. However, according to the model predictions, their findings do not conflict with our results. As discussed in \citet{georgy2019disappearance}, the stars in the mass range of 1.3--1.4\,$M_{\odot}$ at LMC metallicity can develop a significant convective envelope in the late MS and experience magnetic braking. The stellar masses at the MSTO and its bottom of NGC\,1846 are roughly within this range. In other words, the stars may be generating a convective envelope to spin down them, and will eventually be magnetically braked.

To summarize, in this work, we have confirmed that the stars in NGC\,6819, a cluster of 2.5 Gyr old, are fully magnetically braked, by examining the morphology of its TO region as well as the direct measurements of the rotational velocities of its TO stars. On the basis of the PARSEC model, combining the current result with those in the literature, we show that the critical mass of magnetic braking at solar metallicity should lie between 1.54--1.60\,$M_{\odot}$. If only Galactic clusters are considered, this range is relaxed to 1.54--1.97\,$M_{\odot}$. The current work also provides an innovative means to study magnetic breaking based on photometric analyses of the cluster eMSTO phenomenon only, rather than requiring spectroscopic measurements of stellar rotation velocities.

\section*{Acknowledgments}
We thank the referee for many useful comments and suggestions. This work is supported by National Key R\&D Program of China No. 2019YFA0405500 and National Natural Science Foundation of China grants 11903027, 11833006, 11973001, and U1731108. Y. H. is supported by the Yunnan University grant C176220100006. C. L. is supported by National Natural Science Foundation of China grants 12073090 and National Key R\&D Program of China No. 2020YFC2201400. We have used data from the European Space Agency (ESA) mission Gaia (https://www.cosmos.esa.int/gaia), processed by the Gaia Data Processing and Analysis Consortium (DPAC, https:// www.cosmos.esa.int/web/gaia/dpac/consortium).

\bibliography{refrences}{}
\bibliographystyle{aasjournal}

\end{document}